\documentclass[
	a4paper, 
	10pt, 
	unnumberedsections, 
	twoside, 
]{LTJournalArticle}

\runninghead{Extension of Dictionary-Based Compression Algorithms for the Quantitative Visualization of Patterns from Log Files} 

\footertext{\textit{This is a preprint; submitted to EuroVA 2023}} 

\title{Extension of Dictionary-Based Compression Algorithms for the Quantitative Visualization of Patterns from Log Files} 

\author{%
	I. Cherepanov\textsuperscript{1}, J.G. Joewono\textsuperscript{1} , A. Kuijper\textsuperscript{1,2},
    J. Kohlhammer\textsuperscript{1,2}
}

\date{\footnotesize\textsuperscript{\textbf{1}}Fraunhofer IGD, Germany\\ \textsuperscript{\textbf{2}}Technische Universität Darmstadt, Germany \\}

\renewcommand{\maketitlehookd}{%
	\begin{abstract}
Many services today massively and continuously produce log files of different and varying formats.
These logs are important since they contain information about the application activities, which is necessary for improvements by analyzing the behavior and maintaining the security and stability of the system.
It is a common practice to store log files in a compressed form to reduce the sheer size of these files. 
A compression algorithm identifies frequent patterns in a log file to remove redundant information.
This work presents an approach to detect frequent patterns in textual data that can be simultaneously registered during the file compression process with low consumption of resources. 
The log file can be visualized with the possibility to explore the extracted patterns using metrics based on such properties as frequency, length and root prefixes of the acquired pattern. 
This allows an analyst to gain the relevant insights more efficiently reducing the need for manual labor-intensive inspection in the log data.
The extension of the implemented dictionary-based compression algorithm has the advantage of recognizing patterns in log files of any format and eliminates the need to manually perform preparation for any preprocessing of log files.
	\end{abstract}
}

\begin{document}

\maketitle 
    
\section{Introduction}
Logging files (logs) are computer-generated text files.
They are produced by specified events of a running application and contain necessary information about the corresponding condition of the system.
Logging is an essential part of the development and monitoring of a running system. 
There are different tasks behind it such as error analysis, detection of system weaknesses, unexpected events, and anomalies \cite{zeng2016computer, he2021survey, STUDIAWAN20191}.
Moreover, logs provide the possibility to analyze the actual content distribution that passes through the system.
Understanding what the data represents might be useful for a number of things.
Distribution knowledge of the data can be used by analysts to see if the data deviates from the standard behavior or whether a trend can be identified \cite{OUSSOUS2018431, AGRAWAL2015708}. 
Since the services provide massive amounts of log records, it is not feasible to analyze them manually.
For this purpose, automated methods are required.
Many of the proposed approaches depend on particular preprocessing steps for the specific types of logs starting from data preparation to parsing for content extraction from the logs \cite{7579781}.
Further, the extracted data can be presented in a structured form and can also be used for other purposes such as classification \cite{STUDIAWAN20191}.

Since the volume of logs increases over time, it is a common practice to apply compression methods. 
There is a range of different methods for data compression \cite{JAYASANKAR2021119}.
For the logs, it is necessary to apply a lossless compression, so that the decoding reliably restores the original log file.
One of them is the so-called dictionary-based compression that involves replacing variable-length substrings with code words \cite{welch}. 

In this paper, we combine a visualization approach with an extension of a lossless dictionary-based compression technique that allows registering the statistics of the patterns in addition to the compression of the log files. 
The resulting dictionary with acquired patterns allows filtering patterns by the metrics we created. 
The metrics are based on the properties of the patterns such as frequency, root prefix, and length.
The visualization approach shows how original log file data can be visually mapped by means of a color scale in order to represent the quantitative attribute of a selected metric.
This allows analysts to better examine the data as the frequent and infrequent patterns in logs are directly emphasized.
Moreover, our approach can be applied to any type of textual data format without data preprocessing.
This is useful given the large variety of applications in existing systems.

\section{Related Work}
The work associated with our approach falls into three categories. First, we address approaches that process log files and extract knowledge from them.
Then, we describe relevant compression methods and finally, refer to visualization techniques for patterns in log files.

It is essential for complex systems to generate log files \cite{kazemzadeh2009reliable}. 
These log files are used for debugging during system operation. Log files record the events, internal states, and data content of the system during runtime. 
Analysts can better understand the system status by examining the logs and diagnosing the system when an error occurs and finding leverage points to improve the system.
This makes knowledge extraction from logs especially important for the decision-making process \cite{djenouri2018extracting}.
Many approaches are based on log parsing methods, for instance with regular expressions, to extract information \cite{xu2009detecting}. 
These approaches require a non-trivial effort to manually create and maintain expression rules. 
The dynamic and diversely growing variety of applications with unstructured logs makes this process challenging.
Thus, many efforts have been directed towards log analysis through the use of data mining techniques for automated log parsing \cite{zhu2019tools, alspaugh2014analyzing}.
The goal of these methods is to transform the unstructured log data into structured data.
These approaches learn patterns from log data and automatically create common event templates.
These are tools such as Drain \cite{he2017drain},  LogMine \cite{hamooni2016logmine} and MoLFI \cite{messaoudi2018search}.
In this work, we also focused on avoiding the need to manually perform preparation for any preprocessing of log files.
Further, our focus is on an extension of an existing dictionary-based compression technique applied to log files.

The field of compression methods is broad and diverse \cite{JAYASANKAR2021119}. 
There are lossy methods that achieve compression through a reduction of information in the original file, which are appropriate for images and audio files.
In contrast, text files can become worthless if even the smallest segment of the file is eliminated. 
Such data should only be compressed using a lossless compression method \cite{10.5555/3050831}.
One of the methods for textual data is based on the so-called dictionary-based method. 
This method replaces strings of symbols and encodes each string as a code using a dictionary. 
The dictionary stores strings of symbols and corresponding codes.
In this paper, we focus on a popular lossless dictionary-based method called LZW (Lempel-Ziv and Welch) compression \cite{welch}.
The LZW compression proceeds iteratively and at each stage, the input characters are aggregated into a sequence until the next character results in a sequence for which there is no code in the dictionary yet. 
The code for the sequence without that character is added to the output and a new code for the aggregated new sequence is added to the dictionary.
This method can be easily extended with another dictionary that counts all discovered patterns and subpatterns, which we adopt for quantitative visualization.

Our goal is to satisfy the user’s task of discovering frequent and less frequent patterns \cite{6634168}. 
We tackle it with expressive visualization in a log file using different colormaps \cite[Chapter~10]{munzner2015visualization}.
Highlighting the elements of a structure with different colors makes the discovery of patterns preattentive, thus significantly improves the search \cite{WU2003617, treisman1986features}.
Adopting the idea, this work utilizes color highlighting of text sections to emphasize relevant information to the user.
At the same time, the functionality of the compression is exploited with the advantage that no data preprocessing is necessary.
The program named gzthermal2 pursues a similar idea \cite{github}.
It was developed to visualize GZIP compression. 
The program compresses data with the GZIP algorithm and then generates
an image representation of how the data was compressed. 
The various color highlights show the contrast between the number of bytes used to store the data segments in the GZIP file format.
We, on the contrary, represent the quantity of discovered distinct patterns in a compressed file.

\section{Approach}

In this section, we focus on our proposed extension of the LZW algorithm and describe the encoding part of it in more detail.
Then we introduce the metrics that we created based on the frequency, length, and root prefix, which allow the user to sort the data from the dictionary and also on which the quantitative coloring of the log files is performed.
Finally. we describe the visualization of the quantitative properties on the original log file.
We developed our approach as a web-based prototype, from which we generated and represented the results in form of snippets from an exemplary log file.

\subsection{Extention of LZW Algorithm}

\begin{figure}[h!]

\label{code:lzw}
\renewcommand\figurename{Listing}
\includegraphics[width=0.5\textwidth]{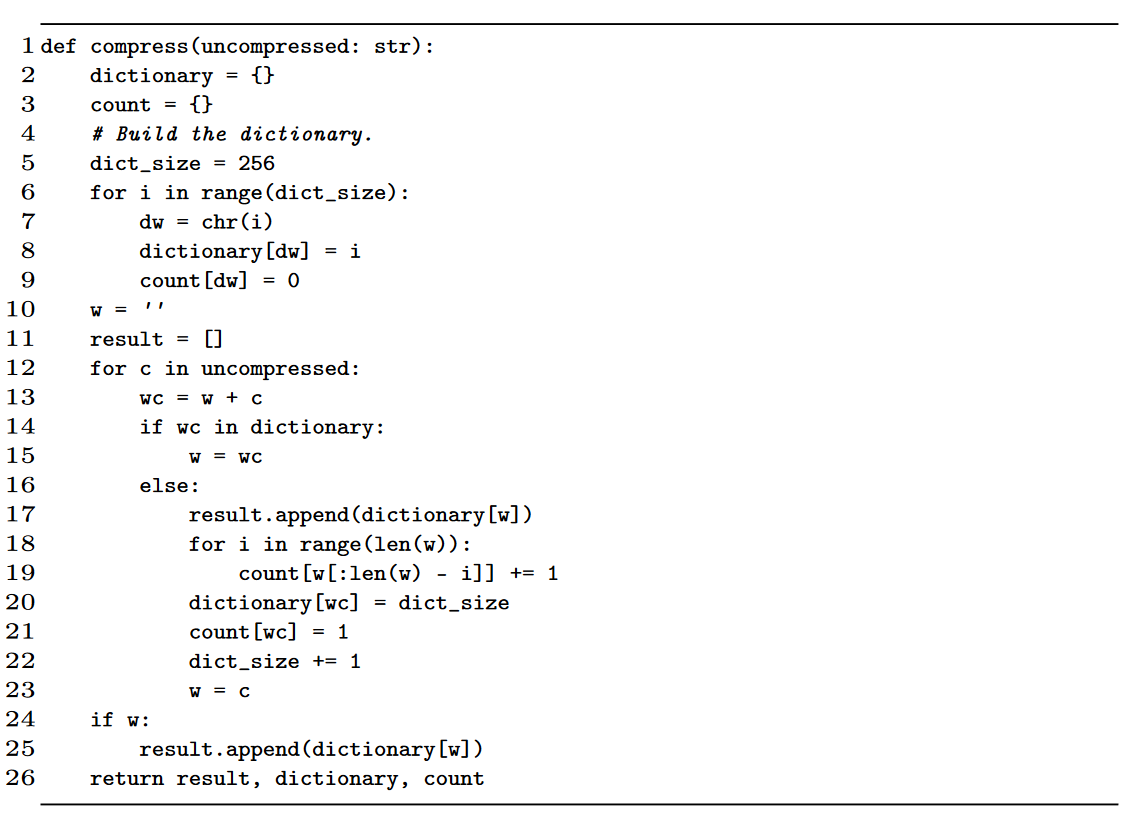}
\caption{Python code of the encoding part of the extended LZW version}
\end{figure}
The first lines of the code initialize two dictionaries. 
One of them stores all patterns with corresponding codes (Line 2) and the other one has the function of a register of pattern occurrences (Line 3).
The LZW method starts with the initialization of the dictionary to all possible symbols in the alphabet (Lines 6-8).
We apply a conversion to ASCII, otherwise, it would have been necessary to initialize the dictionary with all possible Unicode symbols, which increases the overhead.
Since log entries are usually written in the Latin alphabet, the scope of ASCII characters is sufficient for the initial dictionary.
In addition, a dictionary is updated, that records the frequency of all actual and newly occurring patterns and subpatterns, with the same initial patterns (root prefixes) (Line 9).
The encoding method accumulates a symbol to the initial string $w$ (Line 13) in each iteration and checks if the resulting $wc$ is found in the dictionary (Line 12).
If $wc$ is in the dictionary then it is assigned to $w$ (Line 15).
After a certain point, adding the next symbol $c$ causes the search of $wc$ in the dictionary to fail (Line 16).
At this step, the dictionary pointer that points to string $w$ is stored to the encoded resulting array (Line 17) and the string $wc$ is stored as a new dictionary entry with the next available code (Line 20). 
Subsequently, the code increases by one for the next unknown string (Line 22), and since this pattern has been discovered for the first time, it gets assigned 1 for the first occurrence in the dictionary which is responsible for counting patterns (Line 21).
In the same step, we iterate over all subpatterns of the string $w$ and increase their count corresponding occurrence (Lines 18-19).
Finally, string $w$ is initialized to symbol $c$ (Line 23).  
When the input stream is finished we obtain as the result the compressed string, the dictionary of codes with the patterns for later decoding, and the dictionary of counted occurrences of the patterns (Line 26).

The dictionary is not limited by the number of prefixed symbols, the algorithm stores all dictionary entries from the beginning of a file. 
In practice, the dictionary is regularly reset after it becomes too large to allow freeing up memory used by the system and reduce overhead in matching the next bytes in the input stream to the dictionary\cite{welch}. 
The goal of this work, however, is to find as many patterns as possible, so the dictionary should not be reset during the process.
The idea is that the entire log file should be archived and examined for patterns without any splits. 

\begin{figure*}[htb!]
\centering
\subfloat{\frame{\includegraphics[width=0.45\textwidth]{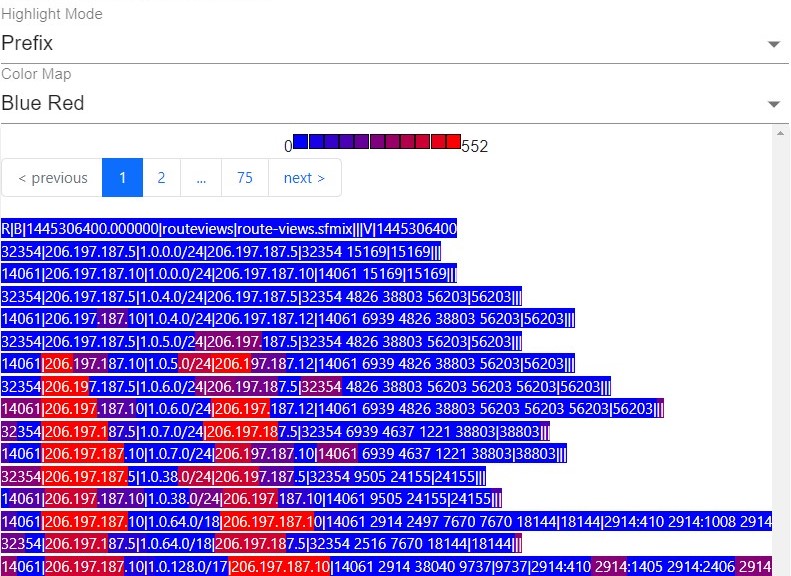}}}\hspace{0.5cm}
\subfloat{\frame{\includegraphics[width=0.45\textwidth]{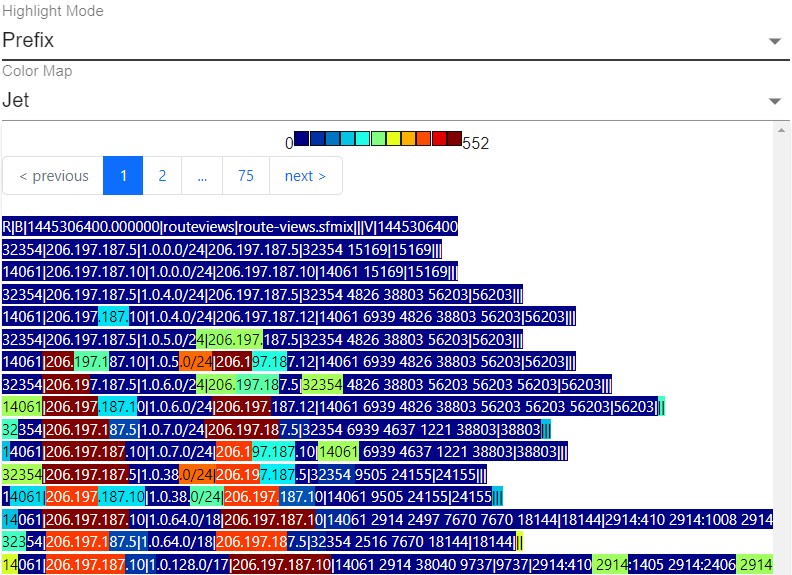}}}
\\
\subfloat{\frame{\includegraphics[width=0.45\textwidth]{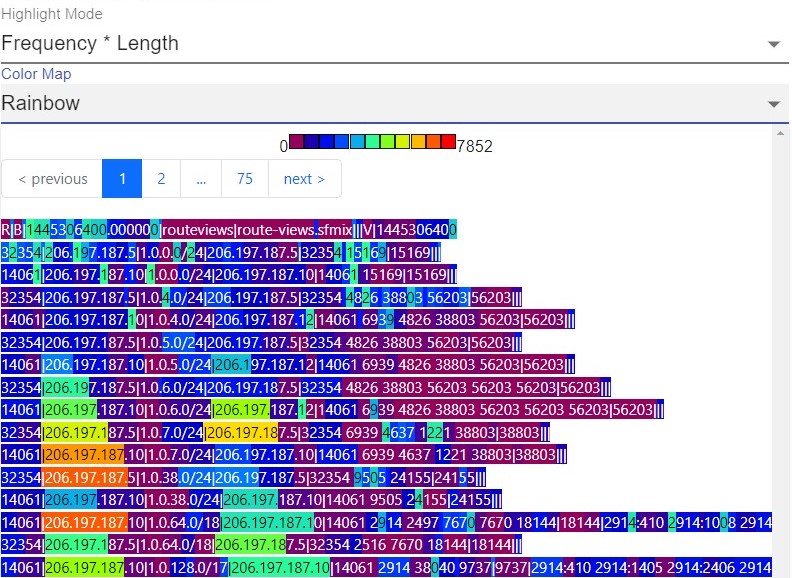}}}\hspace{0.5cm}\subfloat{\frame{\includegraphics[width=0.45\textwidth]{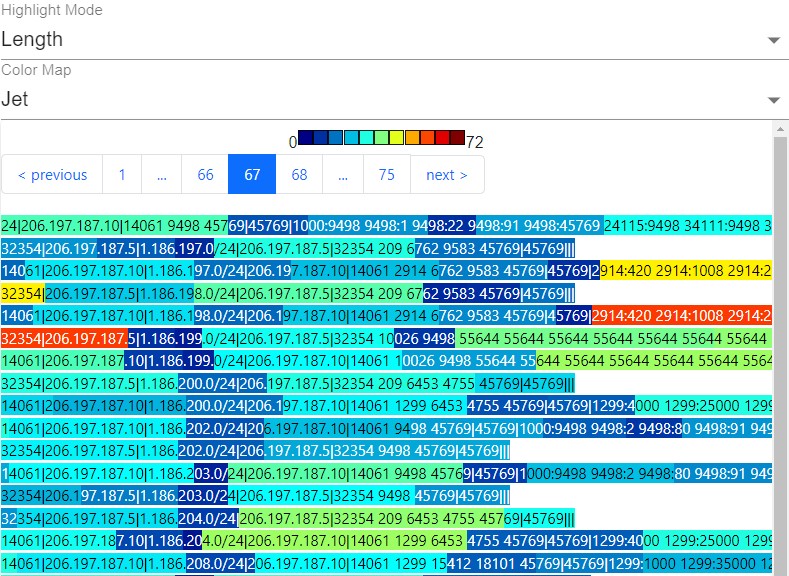}}}\\
\caption{Representation of the colored snippets of a log file. Figure (a) represents the coloring based on the frequencies of the root prefixes of length 5 in a sequential colormap. In contrast, Figure (b) represents the same snippet in jet colormap, which makes the similar quantitative properties more visible in this particular example. (c) illustrates the same snippet with a coloring based on the frequency*length metric. (d) represents a different snippet of the same log file, using the metric for coloring that is based on the length of the pattern.} 
\label{fig:logs}
\end{figure*}

The decoding of the encoded resulting input is performed in the same way as in the original LZW algorithm \cite{welch}.
The decoder goes through the compressed data and uses a code dictionary to translate the codes back to the original text file. 
For later visualization, we append additionally defined quantitative properties to each pattern, which are described in Section \ref{chap:metrics}.
These properties determined the background color of patterns in the presented log visualization as illustrated in Figure \ref{fig:logs}.
The user has the possibility to set the prefix length, which determines the length of the root prefix and set its frequency.
This property can vary and is adjusted if it is changed by the user. 
This is done by searching for the root prefix of the selected length.
In this way, the frequency of this root prefix is obtained and assigned as a pattern property for later visualization.

The resulting modification adds some overhead by having to iterate through the length of each buffer (variable $w$). 
However, for each new pattern stored in the dictionary, the buffer $w$ is reset so that the length of the buffer returns to 1 (Line 23). 
Since the length of the buffer tends toward 1, it can be argued that the overall algorithm has a run-time complexity of $O(n)$.
This is further shown during testing and measuring the average algorithm time. 
The algorithm was tested for the worst-case scenario, where the input consists of repeating the same character $N$ times so that the resulting buffer is reset a minimum number of times and the count is performed on as many subpatterns as possible. 
The results are shown in Table \ref{table:timeeval}. 
The average time is calculated by running the algorithm 20 times in 5 seconds intervals. 
To provide better context for comparison, sample data with regular text with a length of 191,726 bytes takes on average 115.11 ms to process (0.600 ns per $N$). 
A further analysis counted the number of times the counter is accessed and incremented and confirmed that the counter is accessed exactly $N$ times. 
The total running time is approximately $T(2N)$.
\begin{table}[h!]
\centering
\caption{Comparison of average time to process a stream of input consisting of $N$ bytes of the same repeating character.}
\begin{tabular*}{0.49\textwidth}{lrr} 
\toprule
$N$ (bytes) & Average time ($ms$) & Time ($ns$) per $N$\\ 
\midrule 
\tiny 10.000 & \tiny 4.61 & \tiny 0.461 \\
\tiny 20.000 & \tiny 10.17 & \tiny 0.509 \\
\tiny 30.000 & \tiny 14.80 & \tiny 0.493\\ 
\tiny 40.000 & \tiny 20.83 & \tiny 0.521 \\
\tiny 50.000 & \tiny 26.63 & \tiny 0.533 \\
\tiny 80.000 & \tiny 48.55 & \tiny 0.607\\ 
\tiny 160.000 & \tiny 104.04 & \tiny 0.650 \\ 
\tiny 320.000 & \tiny 227.40 & \tiny 0.711 \\
\tiny 640.000 & \tiny 524.06 & \tiny 0.819 \\ 
\bottomrule
\end{tabular*}
\label{table:timeeval}
\end{table}

\subsection{Metrics} \label{chap:metrics}
\begin{figure}[htb!]
    \frame{\includegraphics[width=\linewidth]{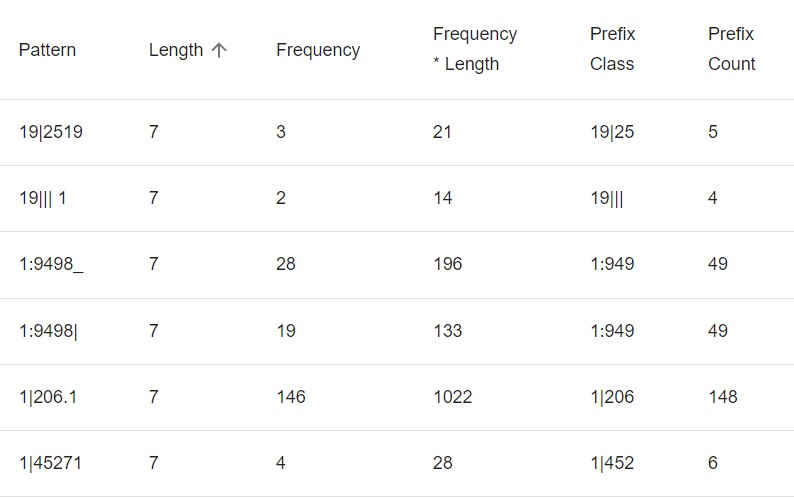}}
    \caption{The tabular overview of the dictionary of acquired patterns.
    Along with the patterns, other metrics are also shown such as the length, frequency, frequency multiplied by length, the prefix, and the frequency of the prefix.}
    \label{fig:table}
\end{figure}

Counting all patterns allows to illustrate them in a table with the possibility to filter the patterns by the following metrics as represented in Figure \ref{fig:table}:
\begin{itemize}[leftmargin=*]
\setlength{\itemsep}{0pt} 
    \item \textbf{Pattern} and its \textbf{frequency} represents all patterns and their frequency. Thus, it is possible to sort by frequency or alphabetically.
    \item The pattern \textbf{length} can be used as a sorting option.
    \item \textbf{Frequency$*$Length}. Frequency multiplied by its length brings out the frequent and long patterns. 
    This metric is useful because it raises more meaningful patterns despite the repetitive meta information that is also present in many other patterns.
    \item \textbf{Prefix class} and its \textbf{count}. The prefix class is an attribute that determines the length of the root prefix from which all longer patterns are derived. 
    The length of the prefixes can be limited by the user to show how many patterns at such length share the same prefix.
    In Figure \ref{fig:table}, the prefix has a length of 5. 
    The frequency of the root prefix is at least equal to the frequency of the corresponding pattern or higher if many patterns follow with the same root prefix.
    All patterns shorter than the set length will be considered rare and  receive a value of 0 for coloring.
\end{itemize}

\subsection{Visualizing Frequent Patterns}

The user's task is to discover frequent and less frequent patterns in a log file.
The frequency is a quantitative value based on various offered metrics (Section \ref{chap:metrics}). These values are represented with various colormaps.
There are several alternatives of colormaps with their strengths and weaknesses, and choosing the right one is difficult \cite{moreland2009diverging}.
For this reason, it is appropriate to offer several colormaps to the user.
The colormap of diverging colors allows us to see if values are near extrema and near which extrema they are located as represented in Figures \ref{fig:logs}a, which is suitable for the separation of very frequent and less frequent patterns.
In this way, frequent patterns can be easily localized by the user due to the color contrast if the entire log file is represented. 
In a narrowed search in a log file, the quantitative values may not differ much and these differences have to be emphasized.
This is when a rainbow similar colormap may become suitable since similar quantitative values have different colors in contrast to divergent coloring as represented in Figures \ref{fig:logs}b, \ref{fig:logs}c and \ref{fig:logs}d. 

The collected patterns are listed in our system in a tabular form  with the following attributes: the pattern itself, the length of the pattern, the frequency, the value of its frequency multiplied by its length, and two more columns showing the number of different patterns which share a prefix with this pattern and the prefix count itself as represented in Figure \ref{fig:table}. 
The user can sort the table based on each attribute.
The patterns can be sorted alphabetically, while the rest are sorted by their nummeric value.
The relevant logs to be examined by an analyst can be decoded into the original file and, depending on the selected metric (Section \ref{chap:metrics}), displayed in color alongside the tabular view with acquired patterns and their frequencies.
This enables the relevant findings to be identified more efficiently, eliminating the need for manual inspection of the entire log data.
Figure \ref{fig:table} represents snippets of the same log file on which different metrics are selected for coloring.
The table and the colored visualization of the original log file might supplement each other.
Since the table can be sorted and directly indicate the frequent patterns and the colored visual view emphasizes the locations in which these patterns are so that an analyst can examine the information next to them.


\section{Usage Scenario}

In order to validate the proposed approach, a case study on a real-world scenario was conducted.
Since saving log files is essential for improving the system, many providers save the logs chronologically in compressed form.
As an example, we are familiar with a provider called The Réseaux IP Européens Network Coordination Centre (RIPE NCC).
The function of RIPE NCC is to act as the Regional Internet Registry providing global internet resources and related services.
It offers services for the benefit of the internet community, including publicly accessible archives of collected Border Gateway Protocol (BGP) update logs.
These update logs are stored chronologically in an archived form with a different method namely gzip and the logs are written in MRT format. 
Based on these logs, there is a number of visualization approaches that can help a domain expert with troubleshooting \cite{9904432}.
These approaches use the provided parser to extract the information from the logs.
Through the exchange with one of the experts of RIPE we validated the idea to supply the compressed archives combined with a helpful representation of the patterns. 

Our approach is also beneficial when the formats of the logs vary in time through adding new applications.
Our approach uses a compression method that can be applied to any textual data type with no need for data preprocessing. 
It is, thus, suitable as a general interface for analysts.
For the dictionary, which contains counted patterns and subpatterns, the tabular overview is helpful (Figure \ref{fig:table}).
A tabular overview allows the analyst to sort the patterns alphabetically but also to sort them by their frequency, their length, or by other criteria (Section \ref{chap:metrics}).
This allows the analyst to see which patterns are common and which are less common. 
This gives the analyst a faster overview of the log file's content, for instance, which IP address is frequently found in the logs.
In addition, the data distribution of the extracted dictionaries can be tracked, so that any deviation in the log events might be directly detected.
The second part is the entire logs that an analyst examines.
With our approach, the patterns are highlighted.
This allows an analyst to detect the relevant insights in a log file more efficiently reducing the need for manual time-consuming inspection of the log data.
The additional metrics allow the user to weigh the properties of the patterns in a different way.
Different colormaps can be used to optimize the perceptible difference between the quantitative properties of the patterns.



\section{Discussion and Future Work}
Our approach avoids the preprocessing or parsing of log data, and can be applied directly to any textual data independent of the syntax format.
However, it is likely that a lot of meta information in the log files is detected as a frequent pattern due to their frequent occurrence.
Much of this redundant meta information is not informative in terms of the content of the log data.
To obtain more useful patterns, prefiltering is required to remove redundant information.
However, the prefiltering process is very specific to each log file format and requires knowledge of the domain.
By simply removing all the repetitive meta information from the logs that are at the beginning of each line, we obtained significantly cleaner patterns. 
This included meta information like the timestamp that is shared by many logs. 
However, it is still possible to browse patterns that are ranked lower.
This was the reason to introduce a new metric that multiplies length and frequency, which raises the more meaningful pattern in spite of the meta information itself.
Thus, we obtained semantically meaningful patterns such as IP blocks and AS routes in the BGP logs \cite{ripe}.
Furthermore, the method was additionally applied to several different text corpora to test its performance in recognizing semantical patterns in literary works. 
The text corpora used in this experiment are from books provided by Project Gutenberg \cite{gerlach2020standardized}.
This showed that the books written in language have the same frequent patterns in congruence with the most frequent stop words \cite{wilbur1992automatic}.  

This paper describes the first version of our approach that we use for the discussion with domain experts to aim at a clear added value over their current process.
In the future, we are interested in exploring other compression approaches to see whether they can also be extended with similiar functionality of pattern extraction and counting.
In particular, such methods that are specially adapted for the log files are interesting as well as the comparison of their performance in terms of pattern extraction \cite{feng2016mlc, lin2015cowic}.
We are going to integrate and extend our approach into existing tools of a domain in combination with other visualization techniques for domain data and are interested in conducting an evaluation with the domain experts to further focus on their requirements with improvements. 
In addition, due to the extensive log files, we want to provide a level-of-detail visualization with zoomable overviews to move between textual details.
The patterns can also be analyzed automatically to choose an adequate colormap based on data heterogeneity.
Moreover, the resulting patterns can be processed with further methods such as classification, similar to methods like Bag-of-Words \cite{9079815}.
The user can also be further involved to interactively create classes by herself.


\section{Conclusion}
In this paper, we have presented a visualization in combination with an extension of a dictionary-based compression algorithm.
This approach allows to count the patterns during compression with the advantage that no preprocessing of the data is necessary.
The method can be directly applied to any textual file regardless of the syntactical format.
We developed a visual concept for the quantitative properties of patterns based on their frequencies and illustrated the results on the exemplary log files.
In addition, we showed that the dictionary can be displayed to the user in a tabular form beside the colored log file.
This allows the user to sort the patterns according to certain properties.
The approach enables the user to fast visually highlight acquired frequent patterns and present their various quantitative properties.

\section*{Acknowledgement}
This research work has been funded by the German Ministry of Education and Research and the Hessian State Ministry for Higher Education, Research and the Arts within their joint support of the National Research Center for Applied Cybersecurity ATHENE.


\printbibliography

\end{document}